# Data stewardship: case studies from North American, Dutch and Finnish universities

Antti Mikael Rousi
*Research Services, Aalto University, Helsinki, Finland*
Reid Isaac Boehm
*Libraries, Purdue University, West Lafayette, Indiana, USA, and*
Yan Wang
*Research Data and Software, Library, Delft University of Technology, Delft, Netherlands*

## Abstract

**Purpose** – As national legislation, federated national services, institutional policies and institutional research service arrangements may differ, data stewardship programs may be organized differently in higher education institutions across the world. This work seeks to elaborate the picture of different data stewardship programs running in different institutional and national research environments.

**Design/methodology/approach** – Utilizing a case study design, this study described three distinct data stewardship programs from Purdue University (United States), Delft Technical University (Netherlands) and Aalto University (Finland). In addition, this work investigated the institutional and national research environments of the programs. The focus was on initiatives led by academic libraries or similar services.

**Findings** – This work demonstrates that data stewardship programs may be organized differently within varying national and institutional contexts. The data stewardship programs varied in terms of roles, organization and funding structures. Furthermore, policies and legislation, organizational structures and national infrastructures differed.

**Research limitations/implications** – The data stewardship programs and their contexts develop, and the descriptions presented in this work should be considered as snapshots.

**Originality/value** – This work broadens the current literature on data stewardship by not only providing detailed descriptions of three distinct data stewardship programs but also highlighting how research environments may affect their organization. We present a summary of key factors in the organization of data stewardship programs.

**Paper type** Research paper

## Introduction

As a way of accelerating scientific discovery, research data sharing has become a part of research and innovation strategies (European Commission, 2019; National Institute of Health, 2020). Data stewards are seen as key in giving guidance to researchers in research data sharing and Findable, Accessible, Interoperable and Reusable (FAIR) data (Boeckhout *et al.*, 2018; Mons, 2018; Wendelborn *et al.*, 2023; also Wilkinson *et al.*, 2016). However, what

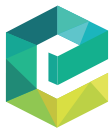

We are grateful for the important commentary provided by Dr Anne Sunikka and Dr Nicole Kong, who helped us to improve the manuscript.

*Conflicts of interest:* The authors are employed by the case study institutions.

constitutes good data stewardship may vary not only per field of science but also upon institutional and national contexts. As national governance, federated national services, institutional service arrangements may differ, data stewardship programs may be organized differently in higher education institutions across the world.

Prior literature has examined data stewardship expertise (Mons, 2018; Wendelborn *et al.*, 2023), stewardship gap areas (York *et al.*, 2018) and data stewardship needs within a specific field of science (Peng *et al.*, 2016; Arend *et al.*, 2022; Awada *et al.*, 2022). There also are initiatives developing data stewardship competencies and profession (Jetten *et al.*, 2021; Wildgaard and Rantasaari, 2022; Whyte *et al.*, 2023). However, the organization of institutional data stewardship programs has not received due attention. The question of how varying national and institutional research environments may affect the organization of institutional data stewardship programs would need further investigation.

This work seeks to elaborate the picture of how different institutional data stewardship programs are organized to fit varying research environments. To this end, three case studies are presented. The included case study universities are Purdue University from the United States, Delft Technical University from the Netherlands and Aalto University from Finland. This work seeks to examine the characteristics of the three case data stewardship programs and to investigate their research environments.

The following terminology is key to the present work. *Data steward* refers to a person(s) working to advise researchers on performing good data management practices or ensure the quality of data assets (Peng *et al.*, 2016; Mons, 2018). *Data stewardship program* refers to the organization of data stewardship activities, including tasks of data stewards, their employment type, their organizational unit and the number of personnel. The present study uses *research data management* (RDM) as an umbrella term for all research data–related services including data stewardship (see Schöpfel *et al.*, 2018). Our analysis of the *institutional and national research environments* relied on the four dimensions of Cox *et al.* (2017), i.e. policy and governance, research data services establishment, funding and structures, advocacy, advisory and support research data services and technical research data services. The key difference between the present work and the study of Cox *et al.* (2017) is that where the previous work focused on the services offered by academic libraries, the present work tries to describe research data service offerings of entire higher education institutions and also the aspect of national research landscape.

The rest of the paper is structured as follows. First, the literature review introduces the prior works pertinent to the topic. Then, the research questions and the data collection and analysis methods of the case study design are specified. The main part of the paper reports the findings from the case studies. The concluding section summarizes the key features of the data stewardship programs and their research environments and reflects the findings.

## Literature review

The characteristics of data stewardship expertise and profession have been investigated in prior literature (Mons, 2018; Wendelborn *et al.*, 2023; Wong, 2023). Additionally, there are initiatives aimed at developing data stewardship competencies, including initiatives to professionalize data stewardship (Jetten *et al.*, 2021; Olapido *et al.*, 2022; Wildgaard and Rantasaari, 2022), the European Open Science Cloud's Data Stewardship, Curricula and Career Paths group (European Open Science Cloud, 2023), and the Skills4EOSC project (Whyte *et al.*, 2023). Whyte *et al.* (2023) provide a useful description of two ends of a data steward role spectrum from Coordinator who is "providing support across research domains" to Embedded who is in a role "close to a research team."

Prior works have also identified data stewardship gap areas (York *et al.*, 2018) and defined data stewardship needs within a specific field of science (Peng *et al.*, 2016; Arend *et al.*, 2022; Awada *et al.*, 2022). Several maturity models concerning research data stewardship activities have also been presented (Peng, 2018), including models for data repository maturity (ISO 16363, 2012; Edmunds *et al.*, 2016), frameworks for assessing data set stewardship maturity (Peng *et al.*, 2015), and data set science maturity (Zhou *et al.*, 2016). Data stewardship in research infrastructures has also been investigated (e.g. Borgman *et al.*, 2019). Although the above works have been important, they do not address the organization of institutional data stewardship programs in higher education institutions.

Studies of research data management service maturity levels often investigate the service models of academic libraries (Pinfield *et al.*, 2014; Cox *et al.*, 2017, 2019; Tenopir *et al.*, 2017; Huang *et al.*, 2021; Singh *et al.*, 2022; Hackett and Kim, 2023). Although the studies of research data service models have been most valuable, they focus on the service offerings academic libraries instead of describing the characteristics of different institutional data stewardship programs and their research environments.

## Research aim
### Research questions
This work seeks to elaborate the picture of how different institutional data stewardship programs are organized to fit varying research environments. First, this work seeks to describe the characteristics of the data stewardship programs running in the case institutions representing North-American, Dutch and Finnish universities.

*RQ1.* What are the characteristics of the data stewardship programs found from case study higher education institutions representing North-American, Dutch and Finnish universities?

Then, this work seeks to investigate the institutional and national research environments of the case programs. Drawing from the work of Cox *et al.* (2017), the research questions for this second task are posed as follows.

*RQ2.1.* What types of policy and governance are affecting the case study institutions?

*RQ2.2.* How are the research data services establishment, funding and structures arranged in the case study institutions?

*RQ2.3.* How are the advocacy, advisory and support research data services arranged in the case study institutions?

*RQ2.4.* What institutional and national technical research data services are most pertinent for the case study institutions?

## Methods
The present study utilized a case study design to answer the research questions. Case study research refers to a broad set of methods used in qualitative inquiries (Swanborn, 2010). In general, case study investigations focus on varying phenomena transpiring within different social systems, such as organizations or local communities (Swanborn, 2010). According to Swanborn (2010), case study design may use several data collection methods.

The study design began with selection of the case study universities; here both variance in data stewardship program types and geographical locations were emphasized. From North-America, Purdue University was selected as the first case study institution. Purdue University was one of the first North-American universities to launch a data management and curation program, including Purdue University Research Repository (PURR) for research data. From continental Europe, Delft Technical University from Netherlands was chosen as the second case study institution as they have a long history of data stewardship services and currently employ full-time data stewards as service personnel at each faculty. From the Nordic countries, Aalto University, Finland was chosen as the third case study because Aalto was one of the first institutions in Finland to launch a data stewardship program consisting of researchers working part-time with data stewardship (Sunikka, 2019). All case institutions were either technical universities (Delft) or emphasize natural sciences and engineering (Purdue, Aalto). It is important to note that when defining the main data stewardship programs of the case institutions, the focus of the present study was on initiatives led by academic libraries, or, in Aalto University's case, similar services. Table 1 provides brief descriptions of the case study institutions.

Our case study design utilized both autoethnography and publicly available documents in its data collection. Autoethnography as a term, in which *auto* refers to the author's critical reflection of personal experience and *ethnography* to qualitative research approach focused on rules and norms of cultural groups, has become a common description for research using systematized views of personal experience as a part of a qualitative methodology (Hughes and Pennington, 2017). Given that the research environments between the case study universities differ, complete similarity of examinations was not possible, even desirable, to attain. Thus, it was important to use the authors' personal experience to further define the institutional policies, structures or processes, or policies of main stakeholder organizations key to each case study organization. Acknowledging the iterative and detailed nature of the present analysis, gathering similar information through interviews would have proven time

| Institution | Country | Brief description | Number of personnel | Number of students |
|---|---|---|---|---|
| Purdue University | United States | Public Land Grant Institution in West Lafayette, Indiana, founded in 1896. It is renowned for engineering and a focus on federally funded research and industry. Purdue has 13 colleges each with multiple departments. Subjects range from STEM to Liberal Arts, to Agriculture, and include a school of Exploratory Studies | Personnel 2022: 10,056 (Faculty 1,978; Staff 8,078) | Students population 2022: 41,000 (Undergrad 75%; Grad 25%) |
| Delft Technical University | The Netherlands | Founded in 1842, oldest and largest technical university in The Netherlands. It has 8 faculties across science, engineering and design, as well as social science and humanities groups embedded in these faculties | Personnel 2022: 6,648 of which 62% scientific staff | Student population 2022: 27,079 (Bachelor and Master's level students) |
| Aalto University | Finland | Formed in 2010 through the merger of Helsinki University of Technology, Helsinki School of Business and University of Arts and Design. Aalto has four schools that focus on technical sciences, one school for business and management and one for arts and design | Personnel 2022: 4,751, 59% of them were employed in teaching and research positions | Student population 2022: 13,000 (Bachelor and Master's level students) |

**Source(s):** Table by authors

Table 1.
Brief descriptions and key figures concerning the case study institutions consuming for interview participants. All the authors have obtained PhDs and managed a team of data stewards at their institutions at the time of writing this work and have worked on average 3.5 years with the data stewardship programs.

According to Swanborn (2010), case studies use several data sources, one of the main ones being publicly available documents. The present study used case publicly available documents to complement its data collection as follows: once the key elements for each of the categories of Cox *et al.* (2017) were defined using autoethnographic methods for the case institutions, pertinent publicly available documents were collected and used both as a basis of further analysis and as means to validate the decisions made in the autoethnographic phase of the research. In summary, the data collection for the case study design consisted of the following main parts.

(1) Selection of case study institutions in August 2022.

(2) The authors wrote autoethnographic case notes of the main features of Cox *et al.*'s (2017) dimensions for their home university, including the organization of the data stewardship programs. The autoethnographic case notes were written during September 2022 to April 2023 and were based on the authors' professional expertise and reflection. The three written autoethnographic case notes comprised a total of 2,369 words and 7 figures and were cross-checked by the other authors to ensure consistency between the cases.

(3) The results from the autoethnographic analyses were presented 10 May 2023 in an open online event for a professional audience. The event received 268 registrations. The autoethnographic case notes were developed based on the feedback of the event participants.

(4) After finalizing the autoethnographic case notes, the authors augmented the data collection with the use of public documents ($n = 45$). The collection of public documents occurred in December 2023. Appendix 1 lists the pertinent public documents per case study institution used in the analysis.

Once the data for the case studies were collected, the authors performed a thematic analysis of the autoethnographic notes and public documents. Thematic analysis is a method for analyzing, identifying and reporting of themes found in data sets such as different textual data (Braun and Clarke, 2006; Nowell *et al.*, 2017). Each of the case study institutions data were first scrutinized to identify the characteristics of the data stewardship programs and the main themes related to the dimensions of Cox *et al.* (2017). Then, the authors examined the main similarities and differences found from the themes linked to each case institution. The findings section that follows presents a synthesis of the full case studies, highlighting both similarities and differences found from the case data stewardship programs and their research environment. The details of all data stewardship programs and their research environments were checked for the last time in December 2023 and may be subject to change.

## Findings
### Data stewardship programs
This investigation revealed two out of three programs where there were full-time data stewards employed in centralized university services.

(1) In Purdue University (Purdue), there were three full-time data stewards employed in the Purdue Libraries organization, and more specifically as research data staff working with the Purdue University Research Data Repository (PURR repository) along with two half-time graduate Assistants dedicated to PURR.

(2) In Delft Technical University (TU Delft), most of the full-time data stewards (11 FTEs) were employed in the faculties' management support teams. In addition, there was one central data steward and one data stewardship coordinator employed in the TU Delft library. Notwithstanding the Purdue's Research Data Department Faculty (see section Research data services establishment, funding and structures), TU Delft has the highest number of FTE assigned to its main data stewardship program.

(3) In Aalto University's (Aalto) program, data stewards were employed mainly in their research departments, often as staff scientists or post-docs. However, a segment of their salary was compensated by the university-level research services to perform data stewardship activities (Darst *et al.*, 2019; Sunikka, 2019). As of 26.5.2023, there were 13 part-time data stewards in Aalto University with total salary compensation of *c.* 1.5 FTEs. Aalto University's data stewards are supported by a fulltime data advisor (1 FTE) located in Research Services that coordinate the network.

In general, the tasks of the data stewards were somewhat similar in all of the case universities. The tasks of the data stewards of all case universities included RDM consultation, support and tooling, RDM training and education, policy and strategy, and community engagement and communications. A notable difference is that the tasks of two of Purdue's stewards were focused specifically on the institutional PURR repository with the repository manager focused on broader data stewardship needs, whereas with TU Delft and Aalto the main data stewardship programs were not arranged around specific services and infrastructures. Table 2 presents a summary of the data stewardship programs, tasks of the stewards and total data stewardship program FTE numbers per institution.

*Policy and governance affecting the case study institutions*
The analysis revealed that all case institutions had legislation and/or policies that focus on access to research and privacy and protection of human subjects. However, some differences could be observed between the North-American (Purdue) and European (TU Delft, Aalto) case study institutions. The North-American policy and legislation landscape had emphasis on privacy, security and responsible conduct of research, whereas European legislation

| Institution | Program type | Tasks | Data stewardship program FTEs |
|---|---|---|---|
| Purdue | Research infrastructure personnel as full-time data stewards<br>Research faculty liaisons – balance tasks with teaching and research | PURR repository management, support in data sharing, archiving and preservation in the repository, PURR repository development, community and campus support department partnerships, RDM consultations and workshop sessions | 3 |
| TU Delft | Full-time RDM professionals employed as data stewards at the faculties management teams | RDM consultation, support and tooling, RDM training and education, policy and strategy, community engagement, communications | 13 |
| Aalto | Researchers as part-time data stewards | RDM consultation, support and tooling, RDM training and education, policy and strategy, community engagement, communications | 2.5 |

**Source(s):** Table by authors

involved aspects regarding access to research, such as the European Open Data Directive (2018). Also the institutional policies of the European case universities addressed access to research. In North-America, access to research was driven by the policies regarding the access to federally funded research (OSTP memorandums on access to federally funded research). In addition, the breadth of national research funders was larger in the United States than in the Netherlands or Finland where the main source of national public research funding was the respective research councils. Table 3 presents a summary of the main legislation and policies affecting the case study institutions.

*Research data services establishment, funding and structures*
The analysis revealed both similarities and differences in the research data services establishment, funding and structures in the case study institutions. In all case institutions, the data steward teams were also supported by full-time library professionals, full-time research software engineers, IT specialists, legal and privacy experts and institutional review boards (named research ethics committees in the European case study institutions). In addition, there were data professionals working in research infrastructures of the case study institutions. Table 4 presents the summary of the main similarities and differences in the research data services establishment, funding and structures of the case study institutions. Notably, Purdue Libraries and School of Information Studies Research Data Department also employed Professors, Researchers and Subject Liaisons who also supported researchers' data handling and sharing.

| Institution | Legislation and policies on access to research | Legislation and policies on privacy and protecting human subjects |
| --- | --- | --- |
| Purdue | Legislation: N/A<br>Policies:<br>Office of Science and Technology Policy Memoranda<br>Increasing Access to the Results of Federally Funded Research (2013);<br>Ensuring Free, Immediate, and Equitable Access to Federally Funded Research (2023) | Legislation: Health Insurance Portability and Accountability Act (HIPAA) (1996); Family Educational Rights and Privacy Act (FERPA) (2021)<br>National Security Presidential Memo - 33 (2021)<br>Policies: Federal Research Misconduct Policy (2000) |
| TU Delft | Legislation: European Open Data Directive (2018); Taverne Amendment (Dutch Copyright Act, 2015) for Research Integrity (2019)<br>Policies: TU Delft Open Access Policy (2016); TU Delft Strategic Framework 2018–2024 (2018); TU Delft Research Data Framework Policy (2018); Dutch Research Council protocol on Research Data Management (2019); TU Delft Research Software Policy and Guidelines (2021) | Legislation: European General Data Protection Regulation (2018)<br>Policies: Netherlands Code of Conduct for Research Integrity (2018) |
| Aalto | Legislation: European Open Data Directive (2018)<br>Policies: Aalto Open Science and Research Policy (2023); Several policies of open science and research in Finland (2019–2023); Academy of Finland Data management and openness policy (2023) | Legislation: European General Data Protection Regulation (2018)<br>Policies: The ethical principles of research with human participants and ethical review in the human sciences in Finland (2019) |

**Source(s):** Table by authors

| Main similarities between case institutions | Main differences between case institutions |
|---|---|
| In all case institutions, data stewardship teams were supported by<br>• Full-time professionals at library services or in units providing similar service functions<br>• Full-time research software engineers<br>• Full-time IT specialists<br>• Full-time professionals at legal services<br>• Institutional review boards (research ethics committees)<br>In all case institutions, there were also data professionals in research infrastructures such as AgData Services (Purdue), 4TU.ResearchData (TU Delft) and Science-IT (Aalto) | Data stewards are employed in different service organizations. In TU Delft, they are faculty staff. In Purdue, they are employed by the library organization. In Aalto, researchers working at the departments receive a part-time compensation from research services for their data stewardship activities<br>Unique to Purdue: Besides PURR research data staff working as data stewards, Purdue Libraries and School of Information Studies Research Data Department also employed Professors, Researchers and Subject Liaisons who also supported researchers' data handling and sharing<br>Unique to Aalto: Aalto has no dedicated library organization, but library service functions reside in research services |

**Source(s):** Table by authors

**Table 4.**
Summary of the
research data services
establishment, funding
and structures in the
case study institutions

*Advocacy, advisory, and support research data services*
In general, the advocacy, advisory, and support topics and methods of delivery were quite similar in all case institutions. The following four themes of similarity emerged from our analysis.

(1) The training topics offered were drawn from the different segments of the data management life cycle. These topics include data management planning, storing of research data, handling of identifiable personal data and research data publishing, for instance.

(2) Legislation and policies on privacy and protecting research participants were key elements of data steward expertise in all case institutions.

(3) This leads to another similarity, which is the capacity to connect and communicate with several actors from other organizational units all at once when resolving support requests.

(4) Lastly, we identified a demand for multimodality of advocacy, advisory and support services (e.g. F-2-F consultations, email support, interactive webinars and recorded webinars). This includes both communications about the data stewardship services and training provided by the data stewards.

Our analysis also revealed some differences in advocacy, advisory, and support research data services. In Aalto, and in some cases at Purdue, the data stewards participated as members in research projects, In Aalto, the data stewards were foremost researchers and their main work was within the research projects. In Purdue, researchers could invite stewards to participate in grants as a co-PI or partner on a case-by-case basis depending on the nature of the collaboration. In TU Delft, the data stewards oversaw the faculty RDM practices and support, but did not join research projects as team members.

*Institutional and national technical research data services*
All case institutions had the following technical services available for researchers: data storing and sharing services during research projects, electronic lab notebooks, software version control systems, and high-performance computing facilities and services. In addition to the institutional services, researchers of all case institutions could also use services of different national infrastructures, such as the National Laboratories and the Agency Data archives in the US, DANS (Data Archiving and Networked Services), SURF (IT cooperation organization of educational and research institutions) and the eScience Centre in the Netherlands, and the Finnish Center for Scientific Computing (CSC), Finnish Social Science Data Archive, and FinData (Social and Health Data Permit Authority) in Finland.

Significant differences could be observed from the in-house repository administration and development occurring in the case study institutions. In Purdue, the data stewardship team administered the institutional PURR research data repository. In TU Delft, a library team – not the data stewards – administered the 4TU research repository services used by a consortium of four technical universities of the Netherlands (TU Delft, Eindhoven University of Technology, University of Twente and Wageningen University and Research). Aalto's services did not run a dedicated repository for research data, but relied on national and European services for data publishing. Table 5 presents a summary of the main similarities and differences between the institutional and national technical research data services.

## Discussion

Drawing from a case study design, this study described the organization of three different data stewardship programs occurring in higher education institutions representing

| Main similarities between case institutions | Main differences between case institutions |
|---|---|
| All case institutions had the following technical services and infrastructures available for researchers<br>• Data storing and sharing services during research projects<br>• Electronic lab notebooks<br>• Software version control systems<br>• High-Performance Computing facilities and services<br>All case institutions had solutions for research information management. Aalto: ACRIS (https://research.aalto.fi/) TU Delft: Research portal (https://research.tudelft.nl/), Purdue has Institutional Data Analytics and Assessment (https://www.purdue.edu/idata/)<br>Symplectic Elements System and use of ORCID identifiers | Differences in in-house data publishing repository development:<br>• Purdue's data stewardship team administered the PURR research data repository;<br>• Delft's Library administered the 4TU research data repository consortium common to the technical universities of Netherlands;<br>• Aalto's services did not run a dedicated repository for research data, but relied on national and European services for data publishing<br>Differences in national services and infrastructures<br>Purdue:<br>• Government agency run labs<br>• Agency Data archives<br>• Not-for-Profit/NGOs<br>TU delft:<br>• DANS: National centre of expertise and repository for research data<br>• SURF: national IT facilities for research and education<br>• eScience Centre: national centre with expertise on research software<br>Aalto:<br>• Finnish Center for Scientific Computing (CSC)<br>• Finnish Social Science Data Archive (based in Tampere University)<br>• FinData: Social and Health Data Permit Authority |

Table 5.
A summary of the main similarities and differences between the institutional and national technical research data services

**Source(s):** Table by authors

North-American, Dutch and Finnish universities. This work broadens the literature on data stewardship (Mons, 2018; York *et al.*, 2018; Peng *et al.*, 2016; Arend *et al.*, 2022; Awada *et al.*, 2022; Wildgaard and Rantasaari, 2022) by providing detailed descriptions of three distinct data stewardship programs and their different research environments. As prior research has focussed in the research data management service-level maturity provided by the academic libraries (Pinfield *et al.*, 2014; Cox *et al.*, 2017, 2019; Tenopir *et al.*, 2017; Huang *et al.*, 2021; Hackett and Kim, 2023), the broader scope of investigation is also a contribution of this work. Appendix 2 summarizes the key characteristics of the data stewardship programs and their research environments.

Our findings align with the valuable definition of data stewardship work presented by Wendelborn *et al.* (2023, pp. 4–5). The present work demonstrates in practice, e.g. the following aspects of their summary: the term data stewardship may be understood differently within different national and institutional contexts; no single person can comprehensively undertake all responsibilities related to data stewardship; building relationships and strategic collaboration that transcends organizational structures requires much effort; and the understanding of the national legal, policy and responsible conduct of research frameworks is key to data stewardship (see Wendelborn *et al.*, 2023, pp. 4–5).

The approach to in-house development of repositories for data publishing paralleled the case institutions size in terms of personnel and student population. In Purdue, with the largest personnel and student population, the in-house developed PURR repository functioned as the general-purpose institutional repository for publishing research data. In the middle-sized TU Delft, the library staff administered the 4TU.ResearchData repository services provided to a consortium of four technical universities of the Netherlands. In the smallest case institution, Aalto, the university services did not run a dedicated general-purpose repository for publishing research data but relied on national and European services for data publishing. This variance in tasks in relation to data repository infrastructures speaks about the breadth of data stewardship expertise, and how this expertise is valuable also without imminent linkage to technical infrastructures. However, in general, the close alliance of technical repository services and domain-expertise of data stewards are often seen as key for further implementation of the FAIR data principles (Dunning *et al.*, 2017; Jacobsen *et al.*, 2020).

TU Delft had the highest number of FTEs in their main data stewardship program, whilst also coordinating with multiple university services in relation to RDM. Although the FTE numbers are an important indicator, the present research highlights that mere FTEs of the main data stewardship program do not fully convey the breadth of research data management service offerings of a higher education institution. An example of this Purdue Libraries and School of Information Studies Research Data Services Department also employed Research Faculty Subject Liaisons and also supported researchers' data handling and sharing. In addition, there were diverse national infrastructures available for the use of the case institutions' researchers, which use is not displayed in the data stewardship programs FTE counts. The development of more sophisticated metrics and cost-benefit analyses of data stewardship programs and other RDM initiatives are important future directions.

In terms of policy and legislation, the differences between the European and the North American regulatory emphasis shown in Table 3 did not seem to lead to significant differences within advocacy, advisory and support research data services provided by the case institutions. Within all institutions, there was much emphasis on privacy and protection of research participants. However, in a scenario where European legislation on research access develops further apart from North American one, then this likely also differentiates the focus of the data stewardship programs between the continents.

Are data stewards in essence service personnel or scientists and which types of career paths they currently have? Our analysis highlights that the data stewards' host organization in an institution has a high impact on their career paths. Researchers working part-time as data stewards have, e.g. field-specific expertise and integration into research groups, and they resume on the career path of researchers. On the other hand, full time data stewards on the service career path can build up a larger resource to meet the increasing demands towards more complex projects impacted by privacy, ethics, legal and research security issues. The authors argue that there is not one solution for hosting data stewards in higher education institutions. Team of full-time data stewards in library or faculty services can be supported by researchers working part-time with RDM, for instance. Interestingly, the library of the largest case study institution, US-based Purdue, had developed into a research performing organization, whereas with the European institutions mergers of service units were visible. Acknowledging the diversity in the organization of the data stewardship programs, the authors agree with the current initiatives to professionalize data stewardship (Jetten *et al.*, 2021; Olapido *et al.*, 2022), as this would help to create sustainable organizations.

In contrast to the differences we have seen in stewardship roles and levels of embeddedness within the stewardship services, we also see elements that lead to greater definition of the data stewardship programs and profession. Whyte *et al.* (2023) describes two ends of a spectrum of Data Steward from Coordinator who is "providing support across research domains" to Embedded who is in a role "close to a research team", which we saw reflected within our analysis. TU Delft can be seen as closer to the Coordinator steward, Aalto to the Embedded steward, and Purdue landing somewhere in the middle. While there is not going to be a clear binary or one specific answer to these questions, continuing to discuss the experiences can bring more insight into the future of the profession. Figure 1 presents the key features found from the research environments of the case study universities that affected the organization of the data stewardship program.

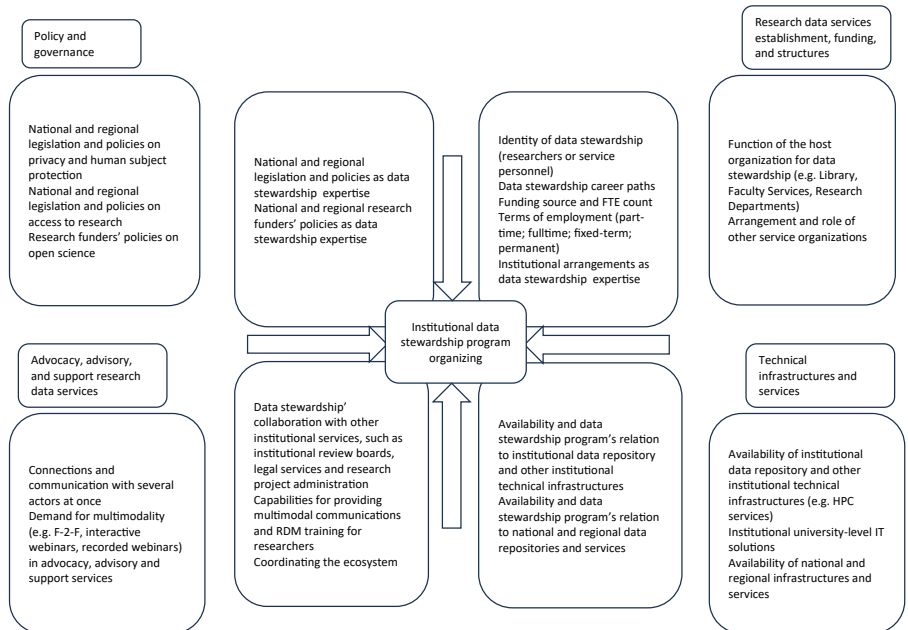

**Source(s):** Figure by authors

We want to highlight the importance of coordinating the data steward team in relation also to the complex institutional and national ecosystems. Building relationships and strategic collaboration that transcends organizational structures is key and requires much effort, which should be duly acknowledged (see also Wendelborn *et al.*, 2023). This applies both to the coordinating of the data steward team and enabling the data stewards to collaborate with the other service units in the higher education institution. Furthermore, we stress the importance of knowledge about the institutional service organizations and national and regional research environments as a key form of data steward expertise.

To summarize, there is no single model for organizing data stewardship programs. Moreover, one needs to consider different institutional and research environments when developing such programs. Drawing from the case studies, we present a 10-point summary of key factors in organizing data stewardship programs. In our view, these factors contribute to how institutional data stewardship transpires. A well-function ecosystem pays attention and seeks quality from the connections of these factors.

(1) Employment type of data stewards (full time vs part-time; fixed-term vs permanent)

(2) Data steward host organization and data steward identity in the higher education institute (e.g. Library, Faculty Services, Research Departments)

(3) Data stewardship program size in FTEs

(4) Data stewardship program's relation to institutional data repository and other institutional technical infrastructures

(5) Leading and coordinating the data stewardship program (hiring data stewards, yearly activity planning, etc.)

(6) Capability for multimodal communications and RDM training for researchers

(7) Collaboration practices with the data stewards and other service units in the higher education institution (e.g. IT and HPC services, Research Software Engineers, Legal Services, Institutional Review Boards)

(8) Data stewardship program's relation to national and regional data repositories and services

(9) National and regional legislation and policies for access to research and privacy and protection of human subjects

(10) National and regional research funders' open science requirements

Lastly, we want to address what we believe our analysis means for academic libraries. As demonstrated by the cases of Purdue and TU Delft, there is no one role for academic libraries, but the role needs to be developed in relation to institutional and national environments. In addition to the RDM capabilities for academic libraries (Cox *et al.*, 2017; Huang *et al.*, 2021; Singh *et al.*, 2022) and fostering the learning of library personnel (Cox *et al.*, 2017, 2019) presented in prior literature, we encourage academic libraries to emphasize building sustainable RDM service organizations and to reserve sufficient resources for coordinating collaboration within their institutional and national ecosystems.

*Conclusions*
This work demonstrates that data stewardship programs may be organized differently within different national and institutional contexts. The study described three different data stewardship programs occurring in higher education institutions representing North

American, Dutch and Finnish universities. The data stewardship programs varied in terms of roles, organization and funding structures. Furthermore, policies and legislation, organizational structures and national infrastructures differed. This work broadens the current literature on data stewardship by not only providing detailed descriptions of three distinct data stewardship programs but also highlighting how the research environments may affect their organization. There is no single model for organizing data stewardship programs. Moreover, one needs to consider different institutional and national research environments when developing such programs. We presented what we believe constitute the key factors of data stewardship program organization. High-quality connections in the core organizational factors are key for a well-performing ecosystem. The development of more sophisticated metrics and cost–benefit analyses of data stewardship programs are important future directions.

*Limitations*
The data stewardship programs and their contexts develop fast, and this research should be seen as a descriptive snapshot taken during a specific time. The autoethnographic methods by design involve subjective elements, and the factors of institutional and national contexts highlighted in the present research should be seen as an interpretation made by the authors with experience in managing data steward teams. Although our summary includes generalizing elements, it is important to note that they are based on only three case studies. Further research would be needed for validation of our results at a more general level (see Eisenhardt, 1989). All of the examined case studies had a focus on technical sciences, which also can be seen as a limitation of this study. Although individual researchers and research groups may undertake data stewardship related activities, this study only investigated formalized services in the case institutions.

## Further reading

## Appendix 1
## List of the pertinent public documents per case study institution that were used in the analysis
### Purdue University

Department of Education (2021), "Family Educational Rights and Privacy Act (FERPA)", Available at: https://www2.ed.gov/policy/gen/guid/fpco/ferpa/index.html (Checked 4 December 2023)

Department of Energy (2023), "National Laboratories", Available at: https://www.usa.gov/agencies/national-laboratories (Checked 4 December 2023)

Department of Health and Human Services (2022), "Summary of the HIPAA Privacy Rule", Available at: https://www.hhs.gov/hipaa/for-professionals/privacy/laws-regulations/index.html (Checked 4 December 2023)

Department of Health and Human Services (2000), "Federal Research Misconduct Policy", Available at: https://ori.hhs.gov/federal-research-misconduct-policy (Checked 4 December 2023)

National Aeronautics and Space Administration (2023), "DATA.NASA.GOV: A catalog of publicly available NASA datasets", Available: https://data.nasa.gov/(Checked 4 December 2023)

National Institute of Health (2023), "NIH, Data Management and Sharing Policy", Available at: https://sharing.nih.gov/data-management-and-sharing-policy (Checked 4 December 2023)

Office of Science and Technology Policy (2013), "Increasing Access to the Results of Federally Funded Research", Available at: https://obamawhitehouse.archives.gov/sites/default/files/microsites/ostp/ostp_public_access_memo_2013.pdf (Checked 4 December 2023)

Office of Science and Technology Policy (2023), "Ensuring Free, Immediate, and Equitable Access to Federally Funded Research", Available at: https://www.whitehouse.gov/wp-content/uploads/2022/08/08-2022-OSTP-Public-Access-Memo.pdf (Checked 4 December 2023)

Purdue University (2023), "Advice and answers from the PURR team", Available at: https://purr.purdue.edu/help (Checked 4 December 2023)

Purdue University (2023), "AgData Services", Available at: https://ag.purdue.edu/department/arge/facilities-and-research/ads/ag-data-services.html (Checked 4 December 2023)

Purdue University (2023), "Purdue Institutional Review Board", Available at: https://www.irb.purdue.edu/(Checked 4 December 2023)

Purdue University (2023), "Research data", Available at: https://www.lib.purdue.edu/researchdata (Checked 4 December 2023)

Scholarly Publishing and Academic Resources Coalition (2023), "Policy and Advocacy", Available at: https://sparcopen.org/what-we-do/active-policy/(Checked 4 December 2023)

**Appendix 2**

| | | Data stewardship program (main program FTEs) | Policy and governance | Research data services establishment, funding and structures | Advocacy, advisory, and support research data services | Technical research data services |
|---|---|---|---|---|---|---|
| | Purdue | Research infrastructure personnel as full-time data stewards (3 FTEs) | Majority Federal Agency based Focus on Security and RCR | Dept: Faculty (Professors, Researchers and Subject Liaisons) and Staff (PURR) Staff leads cross campus focus | Across all data processes Campus connectors | Univ: Storage, sharing, and computing. PURR repository developed in-house Regional to International Collab options |
| | TU Delft | Full-time data stewards employed at the faculties management teams (13 FTEs) | Close connections between institutional and national policies on open access publishing and research data management Comply with EU level legislations on privacy and open data | Research Data and Software services provided by the library Data Stewards employed at faculties | University level advocacy and coordination: led by library research data and software services Research community: led by faculty Data Stewards | Institutional services on RDM with both in-house infrastructure and external applications National infrastructure and facilities coordinated by the library |
| | Aalto | Researchers as part-time data stewards (2.5 FTEs) | The Finnish Federation of Learned Societies coordinates the forming of national open science policies. Institutional policies Comply with EU level legislations on privacy and open data | The data steward programship is funded by research services Research software engineers funded by IT Aalto has no dedicated library organization, but similar functions reside in research services | The data stewards were foremost researchers and their main work was within the research projects Research services coordinate the university-level efforts | Availability of national federated services (e.g. CSC, FinData) and European services Few selected in-house infrastructure initiatives, such as Aalto Materials Database |

**Table A1.**
Key characteristics of the data stewardship programs and their research environments

**Source(s):** Table by authors

**Corresponding author**
Antti Mikael Rousi can be contacted at: antti.m.rousi@aalto.fi